\documentclass[prd,12pt,superscriptaddress,tightenlines,preprintnumbers,showpacs,amsmath,amssymb]{revtex4}

\newcommand{\cS}{{\cal S}}

\newcommand{\cC}{{\cal C}}

\newcommand{\vecn}{{\vec{n}}}

\newcommand{\vecW}{{\vec{W}}}
\newcommand{\vecsigma}{{\vec{\sigma}}}
\newcommand{\vecH}{{\vec{H}}}

\newcommand{\vecX}{{\vec{X}}}

\newcommand{\vecs}{{\vec{s}}}

\newcommand{\bfx}{{{\bf x}}}

\def\bbbone{{\mathchoice {\rm 1\mskip-4mu l} {\rm 1\mskip-4mu l}
{\rm 1\mskip-4.5mu l} {\rm 1\mskip-5mu l}}}
\begin{document}

\title{Spin-Charge Separation and the Pauli Electron}

\author{M. N. Chernodub}\email{Maxim.Chernodub@itep.ru}
\affiliation{ITEP, B. Cheremushkinskaya 25, Moscow 117218, Russia}
\affiliation{Institute for Theoretical Physics, Kanazawa University, Kanazawa 920-1192, Japan}
\author{Antti J. Niemi}\email{Antti.Niemi@teorfys.uu.se}
\affiliation{Department of Theoretical Physics, Uppsala University,
P.O. Box 803, S-75108, Uppsala, Sweden}
\affiliation{Laboratoire de Mathematiques et Physique Theorique
CNRS UMR 6083, Universite de Tours,
Parc de Grandmont, F37200, Tours, France}
\affiliation{Chern Institute of Mathematics, Nankai University,
Tianjin 300071, P.R. China}

\preprint{UUITP-05/06}
\preprint{ITEP-LAT/2006-02}
\preprint{KANAZAWA/2006-03}


\begin{abstract}
The separation between the spin and the charge
converts the quantum mechanical Pauli
Hamiltonian into the Hamiltonian of the non-Abelian Georgi-Glashow model,
notorious for its magnetic monopoles and confinement.
The independent spin and charge
fluctuations both lead to the Faddeev model,
suggesting the existence of a deep duality structure
and indicating that the fundamental carriers of spin and
charge are knotted solitons.
\end{abstract}

\maketitle

Usually, we expect that an electron behaves like
a structureless pointlike elementary particle.
However, recently this view has been challenged
by theoretical proposals \cite{takhta}, \cite{andersson}
that aim to explain
observed phenomena in strongly correlated
environments such as high-temperature
superconducting cuprates \cite{andersson}, \cite{lee}
and fractional quantum Hall systems~\cite{fqhe}.
According to these proposals,
when an electron propagates in isolation or
interacts only with a very low density environment its
spin and charge remain confined into each
other. But in a dense material environment the
strong many-body correlations between
different electrons may force the spin and the charge
to start acting independently.

If present, a spin-charge separation
could have far-reaching practical consequences
to spintronics \cite{spint} that develops devices
which are driven by the spin properties of electrons.
In a wider context~\cite{fad1}, the spin-charge
separation could possibly explain the behavior of elementary
particles in dense environments such as Early Universe
and the interior of compact stars. It might even become
visible in the LHC-ALICE experiment at CERN.

Here we study the spin-charge
separation in the context of non-relativistic spin-$\frac{1}{2}$ particles
that are described by the standard three dimensional,
second quantized Pauli-Maxwell model.
We select this model since its predictions
are likely to lead to experimentally observable consequences in condensed matter
physics. Furthermore, as a field theory model it subsumes structures that
are commonly present in more developed, relativistic quantum field theories.
In obvious notation
\begin{equation}
\mathcal L = \psi^\dagger (i \partial_0 - eA_0 + \mu) \psi
+ \frac{1}{2m}  | (i\partial_k - eA_k)\psi|^2
+ \frac{ge}{2m} \psi^\dagger \vecsigma \cdot
\vecH \psi - \frac{1}{4} F_{\mu\nu}^2\,.
\label{pauli1}
\end{equation}
Here $\psi$ is a two-component commuting spinor,
a Hartree-type wavefunction that describes the nonrelativistic
dynamics of interacting electrons in its
totally antisymmetric subspace. For completeness we have also included
a finite chemical potential $\mu$, and a Zeeman term with
a generic $g$-factor.

We wish to employ (\ref{pauli1}) to describe a quantum state
where the spin and the charge become separated.
For this we choose the wavefunction $\psi$ to describe
a {\it topologically} mixed state where both
the spin-up and the spin-down components are present:
The direction of the spin polarization is a
variable, and specified by a three component unit vector
field $\vecs(\bfx)$. If this vector field approaches a position
independent limit at large distances, $\mathbb R^3$
becomes effectively compactified into a three-sphere. By recalling that
$\pi_3[\mathbb S^2] \simeq \mathbb Z$ we can then employ the nontrivial
homotopy of $\vecs$ to characterize
topological mixing in a spin ensemble, which we build from
the following orthonormal spin-up ($\mathcal X_+$)
and spin-down ($\mathcal X_-$) states
\begin{equation}
\mathcal X_+ = U [\vecs] \left( \begin{matrix} 1 \\ 0
\end{matrix} \right) \ \ \ \ \ {\rm and} \ \ \ \ \mathcal X_-
= U [\vecs] \left( \begin{matrix} 0 \\ 1
\end{matrix} \right) \,.
\end{equation}
Here $U[\vecs]$ is an element of $SU(2)$ and
its relation to  $\vecs$ is
\begin{equation}
\vecs = \mathcal X^\dagger_+\vecsigma \mathcal X^{}_+ =
- \mathcal X^\dagger_- \vecsigma \mathcal X^{}_- \,,
\label{defs}
\end{equation}
so that up to a phase
(\ref{defs}) defines $\mathcal X^{}_\pm$ in terms of
$\vecs$. The spin projection operator
is simply $\frac{1}{2}\vecs \cdot \vecsigma$ and $
\mathcal X^{}_\pm $ are indeed its $\pm \frac{1}{2}$ eigenstates.
When $\phi_\pm$ denote the probability densities
of the spin-up and spin-down components,
\begin{equation}
\psi = \phi^{}_+ \mathcal X^{}_+ + \phi^{}_- \mathcal X^{}_-
= U \left( \begin{matrix} \rho^{}_+ e^{i \Omega_+} \\
\rho^{}_- e^{i \Omega_-}
\end{matrix}\right)
\ = \ \rho U \Phi
\,,
\label{defpsi}
\end{equation}
is our topologically mixed wavefunction.
Here $\rho^{}_\pm$ are the charge densities of the
spin-up and spin-down electrons and $\rho$
is the total density.
We normalize $\psi$ so that the space integral
of $\rho$ coincides with the total number of electrons
in the ensemble. In analogy with (\ref{defs}) we
introduce the three-component unit vector $\vecn =  \Phi^\dagger \vecsigma \Phi$.
In general its $\pi_3[\mathbb S^2] \simeq \mathbb Z$ homotopy class
is similarly nontrivial. As a consequence
our topologically mixed wavefunctions $\psi$ are
classified in terms of the $\pi_3[\mathbb S^2]$ homotopy
classes of {\it both} $\vecs$ and $\vecn$.

We propose that (\ref{defpsi}) entails a separation between the
spin and the charge in $\psi$. Indeed, under a
Maxwellian $U^{}_M(1)$ gauge transformation $\psi \to \exp\{ i\beta\}
\psi$ and $U[\vecs] \in SU(2)$ can not change under this $U^{}_M(1)$
gauge transformation. Instead $\Phi \to \exp\{i\beta\} \Phi$
which identifies the components $\phi^{}_\pm$ as sole carriers
of electric charge.

For the spin, we consider the effect of a global
spatial $SO(3)$ rotation on the spinor $\psi$.
Since the direction of the spin polarization vector
$\vecs(\bfx)$ is variable, we implement a
$SO(3)$ transformation that at
a generic position $\bfx^{}_0 \in \mathbb R^3$
determines a rotation by an angle $\gamma^{}_0$ in
the normal plane of
$\vecs(\bfx^{}_0) = {\vecs}_{0}$,
\begin{equation}
\psi(\bfx^{}_0) \to e^{ \frac{i}{2} \gamma^{}_0\vecs_{0} \cdot
\vecsigma } \psi(\bfx^{}_0)
=  \rho\left[ e^{ \frac{i}{2} \gamma^{}_0\vecs_{0} \cdot \vecsigma}
\cdot U \right] \Phi \,.
\label{rotation}
\end{equation}
For $\gamma^{}_0 = 2\pi$ we indeed have $\psi \to - \psi$
which is consistent with the fermionic nature of $\psi$.  Since
the spatial rotation acts on $U$ by left-multiplication, we
identify $U$ as the carrier of the spin.
This confirms that (\ref{defpsi}) is
a decomposition of $\psi$ into its independent spin and
charge degrees of freedom.

Clearly, the decomposition (\ref{defpsi}) has also a local {\it internal}
$SU(2)$ symmetry that leaves $\psi$ intact but sends $U \to Ug$ and
$\Phi \to g^{-1}\Phi$. This ensures that
both sides of (\ref{defpsi}) describe
four independent field degrees of
freedom. For a given spin polarization direction, the internal
symmetry transformation in general mixes the (relative) probabilities
between the spin-up and spin-down components. Consequently in a given
material environment the local internal $SU(2)$
symmetry must become (spontaneously) broken. Indeed,
the material environment specifies nontrivial ground state expectation
values $\langle\rho^{}_\pm\rangle = \Delta^{}_\pm$. This breaks
spontaneously the internal $SU(2)$ symmetry into a local internal
{\it compact\,} $U^{}_I(1)$ symmetry in the direction of the diagonal Cartan
subalgebra that sends
\begin{equation}
\mathcal X^{}_\pm \to e^{\pm \frac{i}{2} \alpha} \mathcal X^{}_\pm
\ \ \ \ \ {\rm and} \ \ \ \
\phi^{}_\pm \to e^{\mp\frac{i}{2} \alpha } \phi^{}_\pm\,.
\label{intu1}
\end{equation}
This corresponds to (opposite) simultaneous local
rotations in the normal planes of the two unit
vectors $\vecs$ and $\vecn$ respectively.
As a consequence we have a local $U^{}_{M}(1) \times U^{}_{I}(1)$
gauge symmetry; The $U^{}_{I}(1)$ gauge transformation~(\ref{intu1})
when applied only to the charges $\phi_\pm$
provokes a {\it spatial} rotation~(\ref{rotation}) in the normal plane
of the spin polarization direction at the angle $\gamma_0 = - \alpha$.

Note that in terms of the spin variables the internal $SU(2)$ gauge symmetry
determines a {\it local} rotation of the spin quantization axis $\vecs(x)$.
When the $\pi_3[\mathbb S^2]$ homotopy
class of $\vecs$ is nontrivial,
a unitary gauge condition that attempts to {\it
globally} align the direction of the
spin quantization axis {\it e.g.} with
$\vecs(x) \equiv {\bf\hat z}$, overlooks the presence
of topological defects in the homotopically nontrivial $\vecs(x)$.

We substitute (\ref{defpsi}) into (\ref{pauli1}) and
obtain the following {\it remarkable} result:
\begin{eqnarray}
\mathcal L & = &
\frac{1}{2m} (\partial_k \rho)^2
+ \rho^2 (J_0+\mu)
+ \frac{\rho^2}{2m} J_k^2
+ \frac{\rho^2}{8m} (D_k \vecn)^2 \nonumber\\
& & + \frac{1}{16 e^2} \left[\mathcal F_{\mu\nu} -
4 \pi\left(\frac{\rho_+^2}{\rho^2} \widetilde \Sigma^+_{\mu\nu}
+ \frac{\rho_-^2}{\rho^2} \widetilde \Sigma^-_{\mu\nu}\right)
\right]^2\!\!
+ \frac{eg \rho^2}{4m} (\vecH \!\cdot
\overleftrightarrow{\mathbf M} \!\cdot \vecn)
\,.
\label{Pauli2}
\end{eqnarray}
Here $D_\mu = \partial_\mu + \vecX \times $ is the SO(3) covariant derivative,
\begin{equation}
\vecX_\mu  =  \vecW_\mu - 2 J_\mu \vecn\,, \qquad
\frac{1}{2} \vecW_\mu \cdot \vecsigma = i U^\dagger \partial_\mu U\,, \qquad
J_\mu   = - eA_\mu + i \Phi^{\dagger} \partial_\mu \Phi
+ \frac{1}{2} \vecn \cdot \vecW_\mu\,,
\label{defWJX}
\end{equation}
and with ${\vec G}_{\mu\nu} = [D_\mu , D_\nu]$ the $SO(3)$
field strength tensor
\begin{equation}
\mathcal F_{\mu\nu}(\vecX,\vecn) = {\vec G}_{\mu\nu} \cdot \vecn
- \vecn \cdot D_\mu \vecn \times D_\nu \vecn
\label{eq:thooft}
\end{equation}
is the 't~Hooft
tensor \cite{ref:thooft} with $\vecH$ its
magnetic part (that can also accommodate an external
background field), and $\overleftrightarrow{\mathbf M} =
\frac{1}{2} Tr[ U^\dagger {\overleftarrow{\sigma}}
U {\overrightarrow{\sigma}}]$
is the spin quantization frame. Finally, $\Sigma^\pm$ describe the worldsheets
of closed Abrikosov vortices~\cite{orland}
\begin{equation}
\tilde \Sigma^\pm_{\mu\nu} = \frac{1}{2\pi} \partial_{[\mu,} \partial_{\nu]} \Omega_{\pm}
\qquad \mbox{with} \qquad
\Sigma_{\mu\nu} = \frac{1}{2} \epsilon_{\mu\nu\alpha\beta} \widetilde \Sigma_{\alpha\beta}\,.
\end{equation}

Curiously  (\ref{Pauli2}) is essentially
the $SO(3)$ Georgi-Glashow model \cite{ref:thooft} for the multiplet $(\vecn,
\vecX_\mu)$: The material background $\langle\rho^{}_\pm\rangle
= \Delta^\pm$ breaks the $SO(3)$ symmetry spontaneously
into the $U^{}_{I}(1)$ symmetry. This leaves
the  Cartan  $\vecn \cdot
\vecX_\mu$ as the sole propagating component of
$\vecX_\mu$. The local $U^{}_{I}(1)$ gauge symmetry
is spontaneously broken by the mass gap
for (spatial) $J_\mu$, the $U^{}_{M}(1)$ and $SO(3)$
invariant supercurrent that describes the gauge
invariant content of $\vecn \cdot \vecX_\mu$.
The unit vector $\vecn$ is also a propagating degree of
freedom, acquiring a mass gap from the Zeeman term.
Finally $\rho$ too propagates, as a canonical variable
with a conjugate variable that lurks in $J_0$. Thus
(\ref{Pauli2}) describes six independent physical
degrees of freedom, conforming with the four variables
in $\psi$ and the two transverse polarizations in
the Maxwellian~$A_\mu$. Due to the fourth (covariant
derivative) term in (\ref{Pauli2}) the
off-diagonal components of $\vecX_\mu$ are gapped
and non-propagating.

We first consider (\ref{Pauli2})
in a uniform spin background, {\it e.g.} with
$\vecs={\hat{\mathbf z}}$ and $U = \bbbone$,
and in the absence of the Abrikosov vortices. Thus
we can set $\vecW_\mu = 0$ and $\Omega^\pm = 0$.
Since the supercurrent $J_k$ is subject to the Meissner effect
it can be overlooked in the infrared and we obtain from (\ref{Pauli2}),
(\ref{eq:thooft}) the Lagrangian for the
charge degrees of freedom
\begin{equation}
\mathcal L_{\mathrm{charge}}
= \frac{1}{2m} (\partial_k\rho)^2 +
\frac{\rho^2}{8m} (\partial_k\vecn)^2
+ \frac{1}{16e^2} (
\vecn \cdot \partial_\mu \vecn \times \partial_\nu \vecn)^2
+ \frac{ge \rho^2}{4m} (\vecH \cdot \vecn)\,.
\label{fadd1}
\end{equation}
This is the Faddeev model \cite{ludvig} in
interaction with the scalar
field $\rho$, known to support stable knotted
solitons with a self-linking number that coincides with the $\pi_3[\mathbb
S^2]$ homotopy class of $\vecn$ \cite{nature}.  This
suggests that in the uniform spin limit
the elementary excitations in (\ref{Pauli2})
are knotted solitons of charge.

Curiously, we find that knotted solitons
also describe the spin
excitations. For this we consider a uniform
charge distribution represented by a constant $\vecn$
which we align with the positive $z$-axis.
We again look at large distance scales where the Meissner effect
allows us to discard the supercurrent contribution. We then obtain
from (\ref{defs}), (\ref{Pauli2}), (\ref{defWJX}) the
structure (\ref{fadd1}), with $\vecn$ replaced by $\vecs$.

The appearance of the Faddeev model both in the uniform
spin and the uniform charge limits of (\ref{Pauli2})
suggests that in the general case
it describes the interactive dynamics of
knotted solitons of spin and charge. In particular,
we have a manifest spin-charge duality.
Furthermore, since (\ref{Pauli2}) is a descendant of
the Pauli-Maxwell Lagrangian we also have a strong-weak
coupling duality between the Landau-Fermi description
of electron liquid and a description in terms of
knotted spin and charge solitons: Since (\ref{pauli1})
lacks a kinetic term for a $U^{}_I(1)$ gauge field, we can formally
view it as the strong coupling limit. A compact
$U(1)$ gauge theory is a confining theory, with a first-order
deconfinement transition~\cite{Vettorazzo:2004cr}. Since it is natural
to expect that the
coupling in a compact $U(1)$ theory increases with increasing
energy, this explains why at very high energies (and low
densities) the electron behaves like a pointlike particle despite its
nontrivial internal structure. Similarly it explains why at low
energies and/or in proper finite density environments the $U^{}_I(1)$
coupling can become weak leading to a deconfinement of
the spin and the charge with the ensuing decomposition of the electron
into its constituents.

Since (\ref{Pauli2}) relates to the $SO(3)$
Georgi-Glashow model, we expect that it supports a version of the
't~Hooft-Polyakov magnetic monopole. But since the additional
$U^{}_I(1)$ symmetry breaking gaps the supercurrent $J_k$
subjecting it to a Meissner effect, we arrive at the
following proposal: As usual, the
't~Hooft-Polyakov monopole appears as singularity in the
$U_I(1)$ connection $\vecn \cdot \vecX_k$. But
since the off-diagonal components of $\vecX_\mu$ do
not propagate,
the corresponding (bare) off-diagonal correlation length
$\xi_{\mathrm{off}}$ is infinite, and
the monopole has a point-like non-Abelian core.
Due to a Meissner effect in $J_k$ the flux of the diagonal
component is squeezed into two Abrikosov vortices $\Sigma^\pm$.
This confines monopoles and anti-monopoles into magnetically
neutral {\it monopolium} pairs.
Indeed, the topological
content of (\ref{Pauli2}) coincides with
that of compact Abelian Higgs model with two
condensed Higgs fields. The vortices appear as singularities in
the up and down components of the Pauli electron while
non-trivial hedgehog-like spin configurations
corresponds to
the monopoles. In particular,
at the core of vortex $\Sigma^+$ {\it resp.} vortex $\Sigma^-$
the $\rho_+$ {\it resp.} $\rho_-$ spin component vanishes while
at the monopole core $\rho_+ = \rho_- =0$.

Eventually, the off-diagonal component of $\vecX$
may acquire a {\it finite} mass proportional
to $1/ \xi_{\mathrm{off}}$. This may happen
due to quantum corrections, {\it e.g.} in analogy with
high-$T_c$ superconductors~\cite{andersson}. The off-diagonal
$\vecX$ becomes then a propagating degree of freedom.
In elementary particle physics such massive bosons are
common, see {\it e.g.} ~\cite{ref:Greensite}:
When the Georgi-Glashow model is remote from its compact U(1) limit
the magnetic flux of a monopole becomes organized into two
vortices, each carrying half of the total monopole flux.
The ensuing structures are similar to the center vortices
in non-Abelian gauge models~\cite{ref:review:Greensite}
and quite different from the present $\Sigma^\pm$ vortices.
Center vortices have a tendency to organize the Abelian
monopoles into dipole-like and chain-like structures,
which are also present in
Abelian models with doubly charged
matter fields~\cite{Chernodub:2004em}.

The representation (\ref{Pauli2}) suggests that
the spin-up and spin-down components
may condense independently. In particular, in general
the presence of nontrivial condensates
$\langle\rho_{\pm}\rangle \neq 0$ does not
ensure that $\langle\rho_{\pm} e^{i \Omega_\pm}\rangle \neq 0$.
Suppose that there is indeed a state where both components of spin
become frustrated so that $\langle\rho_{\pm} e^{i \Omega_\pm}\rangle = 0$,
while $\xi_{\mathrm{off}}$ remains finite.
The vortices described by $\Sigma^\pm$
are then light, while the center vortices are
heavy. As a consequence monopole dynamics is governed by the
center vortices rather than by the $\Sigma^\pm$ vortices.
In such a state it may then become possible to observe
the monopole-anti-monopole chains proposed in
\cite{ref:Greensite} and \cite{Chernodub:2004em}: a
monopole is a point defect, where
the flux of the vortex alternates.

There may also be a state where only one of the spin
components condenses. Such a {\it partial}
pseudogap phase is then a spin analog of the metallic/electronic
superfluid phase in liquid hydrogen~\cite{egornature}.

The Zeeman term in (\ref{Pauli2}) yields a Josephson coupling between
$\phi_\pm$ which allows {\it e.g.} for a frustration
to spread between the spin components. Note that
the Josephson coupling is absent exactly at points
where $\vecs$ and $\vecH$ are parallel. By applying space-varying
external magnetic field one can emulate phenomena
familiar from physics of Josephson junctions. Both
the partial pseudogap phase and the Josephson junctions
between up-spin and down-spin components are
observable consequences of the spin-charge separation.

In the Georgi-Glashow representation~(\ref{Pauli2}) confinement
leads to area law in the expectation values of the non-Abelian and
Abelian Wilson loops for $\vecX_\mu$ and its diagonal component, respectively.
The Abelian loop
\begin{equation}
W_\cC = \exp\Bigl\{i \int_{\cS} d^2 s_{\mu\nu} \, \mathcal F_{\mu\nu}(\vecX,\vecn)\Bigr\}
      = \exp\Bigl\{i e \int_{\cC} d^2 x_\mu \, A_\mu\Bigr\} \cdot \exp\Bigl\{- i S_{WZ}(\vecn)\Bigr\}\,,
\label{eq:W}
\end{equation}
factorizes into contributions from the Maxwellian and charge fields, respectively.
Since the former does not confine, confinement must manifest itself as a disorder
in the latter. This leads to large values is the Wess-Zumino action,
\begin{equation}
S_{WZ} = \int_{\cS} d^2 s_{\mu\nu} \, \vecn \cdot \partial_\mu \vecn \times \partial_\nu \vecn\,,
\end{equation}
and to the ensuing rapid, area-like
decay of the Wilson loop~(\ref{eq:W}). The disorder can originate
both from the monopoles~\cite{ref:thooft} and the
vortices~\cite{ref:review:Greensite}.

The limit of very strong external magnetic field
incorporates the quantum Hall effect. For this, take $\vecH$ to
contain an external magnetic background component aligned
with the positive $z$-axis. We also restrict
the dynamics into two spatial dimensions by
suppressing fluctuations into the $z$-direction.
In the limit of a very strong external field we then
obtain ${\bf\hat z} \cdot
\overleftrightarrow{\mathbf M} \cdot \vecn \to -1$.
Thus $\vecn = \vecn_0(\vecs) = -
{\bf\hat z} \cdot \overleftrightarrow{\mathbf M}(\vecs)$
which relates the charge variable $\vecn$ to
the spin variable $\vecs$ and since $D_k \vecn_0 \equiv 0$,
\begin{equation}
{\mathcal F}_{\mu\nu}(\vecW,\vecn_0) =
2 \pi \frac{\rho_+^2-\rho_-^2}{\rho^2}
\left(\widetilde \Sigma^+_{\mu\nu}-\widetilde \Sigma^-_{\mu\nu}\right)
+ 2 \pi \widetilde \Sigma^s_{\mu\nu}\,,
\end{equation}
where the spin vortex $\widetilde \Sigma^s_{\mu\nu} = \frac{1}{2
\pi}[\partial_\mu,\partial_\nu] \omega$ corresponds to a
two-dimensional hedgehog $s_1+ i s_2 \propto e^{i \omega}$,
in the shadow of the polarization axis $\vecs$ at the plane which is
perpendicular to the magnetic field $\vecH \propto {\bf\hat z}$. Indeed, we
expect that in general topologically nontrivial
spin vortices are present. But since
the $\vecn \cdot \partial_\mu \vecn \times \partial_\nu \vecn$
contribution to  the 't~Hooft tensor
$\mathcal F_{\mu\nu}$ is absent in the strong field limit,
the magnetic monopoles disappear and (\ref{Pauli2})
reduces to
\begin{equation}
{\mathcal L} \to \frac{1}{2m} (\partial_k \rho)^2
+ \rho^2 (J_0 + \mu) + \frac{\rho^2}{2m} J_k^2
+ \frac{1}{4e^2} \left[ \partial_\mu J_\nu - \partial_\nu J_\mu
+ \pi
\left({\widetilde\Sigma}^+_{\mu\nu}
+ {\widetilde\Sigma}^-_{\mu\nu}
- \widetilde \Sigma^s_{\mu\nu}\right) \right]^2\,.
\label{2d}
\end{equation}
Since $\pi_3[\mathbb S^2] \sim \pi_2[\mathbb S^2]$ our
topologically mixed states  persist in
two dimensions, and assuming that
the Abrikosov vortices become aligned with the $z$-axis we can
replace each of the
string terms  $\Sigma^{\pm}$
by the field strength of an Abelian Chern-Simons.
After averaging over the Gaussian field $J_k$ we find
that (\ref{2d}) reproduces the familiar anyon
approach to  fractional quantum Hall effect including the
Chern-Simons description of its fractional statistics \cite{fqhe}.
In three dimensional the quasiparticles are then knotted
solitons of spin and charge, tightly entangled around
the (closed) Abrikosov vortices that describe
the three dimensional fractional
statistics \cite{omaprl} via
an appropriate generalization of the Chern-Simons action.

The vector field $J_k$ is subject to the Meissner
effect with the ensuing quantization of magnetic flux.
We integrate the supercurrent around a large circle in
the normal plane that surrounds a
vortex configuration, once in a clockwise direction.
We assume an asymptotic London limit where
the electron densities coincide with their constant
background expectation values $\Delta^{}_\pm$. Due to the Meissner effect
the contribution from the supercurrent vanishes and we get for
the magnetic flux \cite{egor2}
\begin{equation}
\oint\! d l_k A_k = - \frac{2\pi}{e}
\frac{1}{\Delta^2} \sum_{i=\pm} \Delta_i^2 N_i
+ \frac{1}{2e} \oint \! d l_k \vecn \cdot \vecW_k\,.
\label{Meissner}
\end{equation}
Here $N^{}_{\pm}$ are the circulations of the Abrikosov vortices
in the spin-up and spin-down components of $\Phi$.
For a finite energy configuration we conclude
from (\ref{pauli1}), (\ref{Pauli2})
that for a partially polarized state  $\Delta_\pm
\neq 0$ and we have {\it (i)} the constraint
$N_+ = N_-$ so that the 1$^{st}$ term
in (\ref{Meissner}) is $2 \pi/e$ times an integer;
{\it (ii)} $D_k \vecn =0$ which implies a relation
between spin and charge variables in the 2$^{nd}$ term,
that is
\begin{equation}
W^a_k \equiv - \epsilon^{abc} {\mathbf M}_{bn}\partial_k {\mathbf M}_{nc}
= - \epsilon^{abc} n^b \partial_k n^c + \lambda_k n^a\,,
\end{equation}
where $\lambda_k \equiv \vecn \cdot \vecW_k$ is the longitudinal part of $\vecW_k$.
If the spin quantization frame
$\overleftrightarrow{\mathbf M}$ is spatially constant, then $\vecn \cdot
\vecW_k = 0$ and the spin-charge mixing term in
(\ref{Meissner}) vanishes. We then recover the standard
flux quantization even though (\ref{Meissner}) {\it a priori} allows for
an arbitrary flux. Note that the presence of
spin vortices in the last term
of (\ref{2d}) suggests that the spin-charge
mixing term may provide the flux quantization in units of $2 \pi/ (2e)$
even though the electron has charge $e$ \cite{egor2}, and even if the
off-diagonal components of $\vecX$-field are non-propagating with
$\xi_{\mathrm{off}}=0$.

The asymptotic flux quantization does not exclude a
fine local structure of the vortices. Despite global flux quantization,
locally the vortices may split into separate
spin-up and spin-down constituents with {\it a priori} arbitrary
fractional fluxes. Furthermore, the flux obtains
nontrivial contributions both from the differences in the relative
charge densities $ \Delta_\pm $ and from the spin-charge mixing.
The spin-up and spin-down vortices are confined into each other by
a logarithmic potential, in configurations which are subject to
the Abrikosov quantization \cite{egor2,egornature}.
We propose that in general, in a strongly correlated system
the vortices become locally composed of spin-up and spin-down
components. However, we suspect that the vortices with a double-core
structure should be metastable due to the attraction
of their cores.

Finally, the domain walls that connect the spin vortices described by $\Sigma^s$
along the lines of the sign flips $\psi \to - \psi$ should confine these
vortices into spatially finite regions.  We expect that the spin-vortex
structures as well as the double-core vortices should
be visible in fractional quantum Hall experiments~\cite{fqhe} and
spintronic devices~\cite{spint}.

A.N. thanks L. Faddeev, and we both thank K. Zarembo  for discussions.
M.Ch. thanks the Departments of Theoretical Physics
of Uppsala
and Kanazawa Universities for kind hospitality.
This work has been supported by a VR Grant and by
STINT Institutional Grant. M.Ch. is also supported by the JSPS grant No. L-06514.

\end{document}